\documentclass[11pt,a4paper]{article}
\usepackage{jcappub}
\usepackage{bm}
\usepackage{bbm}
\usepackage{amsmath}
\usepackage{etoolbox}
\usepackage{float}
\usepackage{mhchem}
\usepackage{gensymb}
\usepackage{color}

\def\circa#1{\,\raise.3ex\hbox{$#1$\kern-.75em\lower1ex\hbox{$\sim$}}\,}

\newcommand{\DM}{{\rm DM}}

\newcommand{\beq}{\begin{equation}}
\newcommand{\eeq}{\end{equation}}

\definecolor{seagreen}{rgb}{0.18, 0.55, 0.34}

\newcommand{\gag}{g_{a\gamma}}

\usepackage{etoolbox}

\makeatletter
\gdef\@fpheader{}
\makeatother

\begin{document}

\makeatletter

\title{Searching for axion-like particles with SPHEREx}

\author[a,b]{Marco Regis,}
\author[b]{Marco Taoso,}
\author[b]{Jorge Terol Calvo}

\affiliation[a]{Dipartimento di Fisica, Universit\`{a} di Torino, via P. Giuria 1, I--10125 Torino, Italy}
\affiliation[b]{Istituto Nazionale di Fisica Nucleare, Sezione di Torino, via P. Giuria 1, I--10125 Torino, Italy}

\emailAdd{jortecal@protonmail.com}
\emailAdd{marco.regis@unito.it}
\emailAdd{marco.taoso@to.infn.it}

\abstract{We study prospects to detect axion-like particles (ALPs) with the upcoming near-infrared telescope SPHEREx.
The signal under investigation is the ALP decay into two photons. Assuming dark matter (DM) to be in the form of ALPs, we analyze the signal from the DM halos of dwarf spheroidal galaxies, the Large Magellanic Cloud and the Milky Way. We find that SPHEREx can significantly improve current limits on the axion-photon coupling in the 0.5-3 eV ALP mass range. }

\maketitle


\section{Introduction}

Axion-like particles (ALPs) stand out as one of the most compelling candidates for Dark Matter (DM) \cite{Preskill:1982cy, Abbott:1982af, Dine:1982ah}. They are light pseudo-scalar particles, generalizing the case of the QCD axion~\cite{Peccei:1977hh,Peccei:1977ur, Weinberg:1977ma, Wilczek:1977pj}, and appearing in various beyond-the-standard-model scenarios~\cite{Arias:2012az}.
ALPs interact with photons through the Lagrangian term $\mathcal{L}=-\frac{1}{4}\gag\,a\,F_{\mu\nu}\tilde{F}_{\mu\nu},$ where $a$ is the ALP field, $F_{\mu\nu}$ is the electromagnetic field strength, $\tilde{F}_{\mu\nu}$ its dual, and $\gag$ the coupling constant.
 This interaction enables ALPs to decay into two photons, each carrying an energy equal to half the ALP mass $m_a$ in the ALP rest frame. In astrophysical environments, dark matter ALPs have low momentum dispersion compared to their mass, and the decay into photons results in a narrow spectral line signature.

Several searches have been conducted for such a spectral line at optical~\cite{Grin:2006aw, Regis:2020fhw, Todarello:2023hdk, Wang:2023imi}, and near-infrared~\cite{Janish:2023kvi, Yin:2024lla} frequencies, yielding exclusion limits on $\gag$ surpassing those from stellar evolution in globular clusters~\cite{Ayala:2014pea, Dolan:2022kul}.

In this work, we explore the capability of the telescope Spectro-Photometer for the History of the Universe, Epoch of Reionization, and Ices Explorer (SPHEREx) \cite{2020SPIE11443E..0IC, SPHEREx:2014bgr,SPHEREx:2016vbo,SPHEREx:2018xfm} in this search.
SPHEREx is a space observatory developed by NASA, and scheduled for launch in late February 2025. It will conduct a two-year mission to map the entire sky in near-infrared light, from 0.75 to 5.0 $\mu$m. 
SPHEREx will provide a step forward with respect to the 2MASS survey~\cite{2MASS}, by extending the wavelength coverage and improving the sensitivity reach. The all-sky spectral survey created by SPHEREx will allow to explore cosmic phenomena spanning from the local regions of our Galaxy to very distant extragalactic sources.
SPHEREx is a wide-field telescope, i.e., if compared to other  satellites operating at similar wavelengths, like the Hubble Space Telescope (HST) and
the James Webb Space Telescope (JWST), it has a very large field of view. This feature makes SPHEREx particularly suited for detecting extended emissions, such as the diffuse emission produced by ALPs in DM halos. 

We will investigate infrared light from the ALP radiative decay in three different targets, dwarf spheroidal galaxies (dSphs), the Large Magellanic Cloud (LMC), and the halo of the Milky Way.
Another potential avenue, which we do not explore here, involves investigating the cosmological ALP signal given by the collection of lines arising in different DM halos, through line intensity mapping~\cite{Creque-Sarbinowski:2018ebl,Bernal:2020lkd}.  These astrophysical probes will be complementary to laboratory searches aiming at the same region of the parameter space such as \cite{Baryakhtar:2018doz}.

The axion decay rate scales as $m_a^3$, making it increasingly challenging to constrain ALPs as one moves toward the infrared regime. Large fields of view (to integrate more signal) and superior flux sensitivities (to enhance statistics) are essential to reach competitive constraints. Here below we show that SPHEREx can surpass existing bounds on $\gag$ for ALP masses between 0.5-3 eV.

The paper is organized as follows. The calculation of the line signal from ALP decay is described in Sec.~\ref{Sec:Lines}. SPHEREx capabilities are outlined in Sec.~\ref{Sec:spherex}. 
We present our results in terms of forecasts for the sensitivity on the axion-photon coupling $\gag$ in Sec.~\ref{Sec:res}. Sec.~\ref{Sec:Conclusion} concludes.

\section{Spectral line signatures from ALP decays}\label{Sec:Lines}

As mentioned in the Introduction, ALPs populating a DM halo can produce a radiative signal from their decay. The energy of the two photons produced in the decay is $E_\gamma \approx \frac{1}{2}m_a \left(1+ \mathcal{O}(v/c)\right)$, where $v$ is the ALP velocity dispersion in the DM halo, about $10^{-3}\, c$ in the Milky Way and lower for the other targets we consider in this work. 

The ALP decay rate is given by 
\beq\label{eq:AxionDecay}
    \Gamma_{a\to \gamma\gamma} = \dfrac{g_{a\gamma}^2 m_a^3}{64\, \pi}.
\eeq
The resulting flux of photons coming from the ALP decay in a nearby source is
\beq\label{eq:AxionFlux}
  \Phi_{\DM} =  \frac{\Gamma_{a\to \gamma\gamma} }{4\pi} D  \quad \quad {\rm with} \quad \quad D\left( \theta_0,\,\Delta\Omega \right) = \int_{\Delta\Omega}{\textrm d}\Omega\, \int\displaylimits_{\rm l.o.s.} {\textrm d} s \; \rho\left( r\left[ s, \theta,\theta_0 \right] \right)\;.
\eeq
Here $D$ denotes the D-factor, which encodes the integral of the DM density along the line-of-sight and across the telescope aperture $\Delta\Omega$, $\theta_0$ is the angle between the direction of observation and the halo center, $\theta$ is the angle between the line-of-sight and the direction of the observational pointing, $r$ is the distance to the halo center, and $s$ is the distance along the line-of-sight. 
The flux described in Eq.~(\ref{eq:AxionFlux}) is integrated over the energy width of the line signal.
Indeed, as mentioned below, we consider spectral channels matching the SPHEREx spectral resolution, which is much larger than the intrinsic width of the photon line induced by the ALP velocity dispersion.

Our reference profile to describe the DM spatial distribution in halos is the Navarro-Frenk-White (NFW) profile~\cite{Navarro:1995iw}:

\beq \rho(r) = \rho_s\dfrac{r_s}{r}\left( 1 + \dfrac{r}{r_s}\right)^{-2}, \eeq
where $r_s$ is the scale radius and $\rho_s$ is the normalization. Their numerical value will be specified for each target in Sec.~\ref{Sec:res}. 
Since we will consider large fields of view, and since the bounds on $\gag$ scales with the square root of the profile, the choice of the parametrization for the DM profile is not very crucial in our study. 

In Eq.~\ref{eq:AxionFlux}, we neglected absorption. Dust extinction can be a relevant absorption process at ultraviolet frequencies. However, SPHEREx operates at the longer wavelengths, where its impact is less severe. Moreover, the focus of our search on the Milky Way is not on regions with very high dust density, as, e.g., towards the Galactic Center~\cite{Roy:2023omw}. The LMC, however, has regions with strong extinction, such as Doradus 30. They are also bright spots, and the strategy we will use for the ALP search includes masking those regions. We estimated dust extinction from Refs.~\cite{2011ApJ...737..103S,2020ApJ...889..179G}, finding an impact on the bounds on $g_{a\gamma}$ below a few percent at 0.75 $\mu$m (the upper edge of the SPHEREx frequency range) and smaller at longer wavelengths. 

\section{SPHEREx capabilities}\label{Sec:spherex}
SPHEREx~\cite{2020SPIE11443E..0IC, SPHEREx:2014bgr,SPHEREx:2016vbo,SPHEREx:2018xfm} is a wide-field satellite telescope composed by two detector arrays, each one providing a field-of-view of $3.5^\circ \times 11.3^\circ$.
The telescope operates at wavelengths between 0.75 $\mu$m and 5 $\mu$m, with a pixel resolution of $6.2"\times 6.2"$.
The spectral resolving power $R=\lambda/\Delta\lambda$ ranges from 40 at 0.75 $\mu$m to 130 at 5 $\mu$m.

The mission is expected to last for two years performing an unprecedented all-sky spectroscopic survey.
The scanning strategy is such that the full sky is covered in 6 months, thus having four surveys in two years.
Each of the four surveys will include two fields that, due to the specific orbit of SPHEREx, are observed for a longer duration (Deep Fields) with respect to the rest of the sky. The area of the Deep Fields is about 100 square degree regions and their location is near the ecliptic poles (we assume they are exactly centered at the north and south ecliptic poles in our computations).

The flux described in Eq.~(\ref{eq:AxionFlux}) is integrated over the ALP energy. Indeed, the ALP velocity dispersion is significantly smaller than the SPHEREx spectral resolution and the ALP emission is thus entirely contained in a given spectral channel of SPHEREx. 
Among the data products, SPHEREx will release spectral images of the sky, and they can be used to perform the search for ALP radiative decays. The predicted surface brightness sensitivity (1$\sigma$) in each pixel ranges from 27 (3.8) nW/m$^2$/sr at 0.75 $\mu$m  to 16 (2.3) nW/m$^2$/sr at 5 $\mu$m for the all-sky (Deep Fields) survey,  for spectral channels matching the FWHM energy resolution, and for the full 2 year nominal mission~\footnote{\href{https://github.com/SPHEREx/Public-products}{https://github.com/SPHEREx/Public-products}}.
These noise levels include both the read noise and the photon noise coming from the Zodiacal light.
They will be used in the next Section to derive the projected sensitivity to ALP signals.

\section{Forecasts} \label{Sec:res}

We estimate the sensitivity of SPHEREx to ALP decays in a given target comparing the DM signal and the surface brightness sensitivity in all the $6.2"$ pixels in a region of interest (ROI) encompassing the target.
Namely, we derive the $95\%\,{\rm C.L}.$ sensitivity reach from the $\chi^2$ (assuming gaussian distributed data we require a threshold value of 3.84):
\beq 
\chi^2 = \sum_i \frac{\Phi_{\DM\,i}^2}{\sigma^2}, 
\label{eq:chi2}
\eeq
where the index $i$ runs over the different pixels, and $\sigma$ is the surface brightness sensitivity in a pixel for the all-sky or Deep Fields surveys, depending on the target under consideration.

With real data at hand, there are two aspects of the analysis that are not discussed here and that will be added to obtain the forecasted sensitivity.
One is background subtraction. The continuum spectrum will be accounted for by fitting a parametric function to data. The second aspect to consider is the masking of bright sources. This can be faced following the approach in Ref.~\cite{Regis:2020fhw} and taking advantage of the source catalog that will be publicly released by the SPHEREx collaboration~\cite{2020SPIE11443E..0IC}.
Considering the expected number of galaxies that SPHEREx will detect~\cite{SPHEREx:2016vbo} (and in particular galaxies with detectable emission lines, see~\cite{Feder:2023rqg}), one can estimate that a limited fraction of the sky shall be masked, and therefore our sensitivities should not be affected dramatically.

\subsection{Dwarf spheroidal galaxies}

dSph galaxies are among the most promising targets for indirect DM searches, given that they are DM-dominated objects and their proximity to us.
We consider the sample of eight dSphs analyzed in Ref.~\cite{Song:2024vdc}.
In each target the DM density is modeled as a NFW profile truncated at distances above the tidal radius, in order to account for the tidal stripping by the Milky Way halo. 
For the ROI we consider a circle around the center of the dSph galaxy up to an angle corresponding to the smallest between the tidal radius and the radius of the outermost star in the object. This angle is specific for each target and it is typically of $\mathcal{O}(1)$ deg. This quantity and the parameters of the density profiles are taken from Ref.~\cite{Song:2024vdc}. 
The D-factor of the Draco dSph, one of our targets, is shown in Fig.~\ref{fig:Dfactors} as a function of the integration angle $\theta$ defined from the center of the object, see Eq.~(\ref{eq:AxionFlux}) with $\Delta\Omega=2\pi\,(1-\cos\theta)$. As evident, when approaching $\theta\sim 1$ deg, ALP decays in the Milky Way halo can also contribute significantly to the signal. 
This emission is spatially uniform on those scales,
whilst the signal from the dSph halo has a characteristic morphology, which, leveraging the excellent angular resolution of SPHEREx, can be exploited to enhance the sensitivity. 
For this reason, here we neglect the contribution of the MW halo to the signal.

None of the dSphs under consideration falls inside the two SPHEREx Deep Fields, therefore the surface brightness noise in Eq.~(\ref{eq:chi2}) is taken to be the one for the all-sky survey.
In Fig.~\ref{fig:FullSensitivity} we present the combined sensitivity reach on the axion-photon coupling $g_{a\gamma}$ obtained by summing the $\chi^2$ of all the eight dSphs. We have found that the reach is dominated by the Draco and Ursa Minor dSphs, whose individual sensitivity is only a factor $\sim 1.4$ lower than the combined one.

\subsection{Large Magellanic Cloud}

The Deep Field survey presents a great opportunity to search for spectral lines from ALP decays given its improved sensitivity with respect to the all-sky survey. One of the two Deep Fields is centered at the south ecliptic pole, which is located at the galactic coordinates $b=-29.81\degree,\, \ell=276.39\degree$. 
Interestingly, the center of the LMC is separated by an angular distance of 4.65 degrees, implying that a significant fraction of the LMC's halo falls inside the 100 square degree of the south Deep Field.

We model the DM density halo in the LMC with a NFW DM profile, with parameters from Ref.~\cite{Regis:2021glv}. We consider a ROI obtained as the intersection between the area of the Deep Field survey and a circular region of 3 degree radius around the LMC center, that comprises a total area of $\sim 18\, \rm{deg}^2$.

Notice that the LMC halo extends at much larger distances. However, the kinematic data used in Ref.~\cite{Regis:2021glv} to extract the DM density profile extend up to $\sim3$ degrees, which means that the density profile beyond that radius is more uncertain. Moreover, we find that the sensitivity reach to the ALP signal does not significantly increase for larger ROIs nor does change significantly if one of the other three DM density profiles studied in \cite{Regis:2021glv} is used.

The surface brightness noise that we use in Eq.~(\ref{eq:chi2})  is the average value inside the Deep Fields mentioned in Sec.~\ref{Sec:spherex}.
This noise level includes the instrumental noise and the noise induced by the subtraction of the zodiacal light emission. 
The intrinsic emission of the LMC constitutes an additional foreground. From measurements of the global continuum spectrum of the LMC we estimate that, in quiet regions, this foreground is below or of the order of the zodiacal light emission, see, e.g., Refs.~\cite{Israel:2010zj,Paradis_2023}. Therefore, masking bright regions in the LMC and with an appropriate cleaning procedure, we expect that most of the large scale LMC emission could be subtracted without a significant increase of the noise level.

The $95\%\,{\rm C.L}.$ projected sensitivity is shown in Fig.~\ref{fig:FullSensitivity}. We find that the larger D-factor of LMC (see Fig.~\ref{fig:Dfactors}) and the improved surface brightness noise of the Deep Field survey enhance considerably the sensitivity with respect to the dSph case, reaching values of the axion-photon coupling $g_{a\gamma}$ below current bounds.

\subsection{Milky Way Halo}

The second Deep Field that will be covered by SPHEREx is located in the north ecliptic pole ($ b = 29.81\degree,\, \ell = 96.38\degree$). In this region, we do not identify any individual nearby object that could be an interesting target for our search. On the other hand, being located at high latitudes, the field presents limited foreground emissions.
This makes the north Deep Field a compelling opportunity in the search for DM lines from ALP decays in the Milky Way halo.

The Galactic ALP signal is almost uniform inside the field, but being nearly monochromatic one can exploit this peculiar spectral property to distinguish it from the dominant continuum astrophysical foregrounds. 
To evaluate the sensitivity on $g_{a\gamma},$ we compute the D-factor corresponding to a $100$ deg$^2$ region centered at the north ecliptic pole,
using the parameters of the NFW Milky Way halo $r_s = 14.46$ kpc and $\rho_s = 0.566$ GeV/cm$^3$ given in Ref.~\cite{Cirelli:2024ssz}.
We obtain a similar D-factor employing the Einasto DM profile reconstructed from kinematical data in Ref.~\cite{Ou:2023adg}.

\begin{figure}[H]
    \centering
    \includegraphics[width=0.75\textwidth]{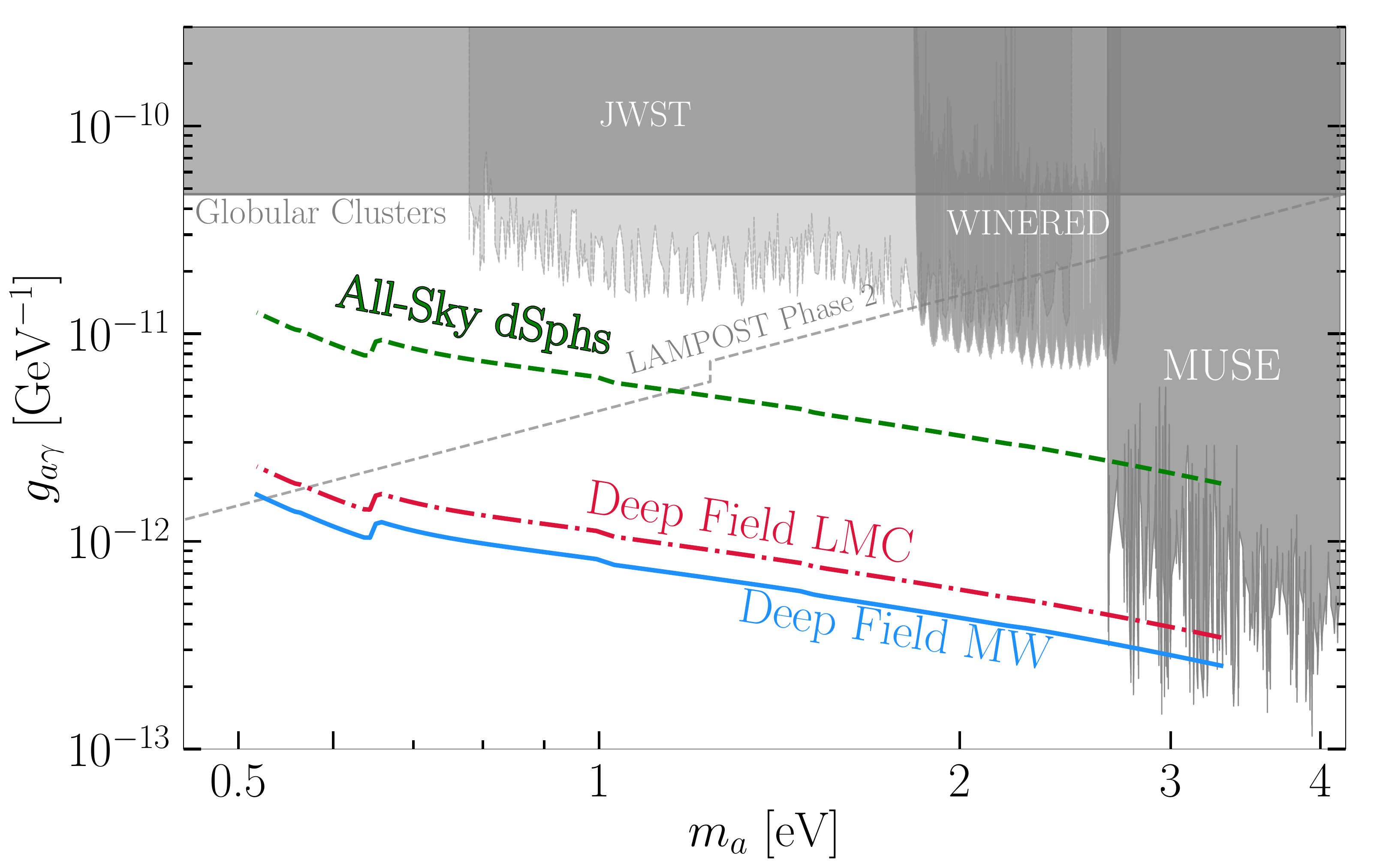}
    \centering
    \caption{$95\%\,{\rm C.L}.$ sensitivity of SPHEREx to the axion-photon coupling $g_{a\gamma}$ for the three targets considered in this work. 
    The dashed green line is from the combined observation of 8 dSphs with the all-sky survey, the dash-dotted red line from the LMC's halo within the south Deep Field, and the solid blue line from the Milky Way halo region observed in the north Deep Field.
    Shaded grey regions are existing constraints from searches of the ALP decay signal with MUSE~\cite{Todarello:2023hdk}, WINERED~\cite{Yin:2024lla} and JWST~\cite{Janish:2023kvi}, and from the impact of ALPs on the stellar evolution in globular clusters~\cite{Dolan:2022kul}. The dashed grey line corresponds to the sensitivity of the LAMPOST Phase 2 experiment \cite{Baryakhtar:2018doz}
    }
    \label{fig:FullSensitivity}
\end{figure}

\begin{figure}[H]
    \centering
    \includegraphics[width=0.75\textwidth]{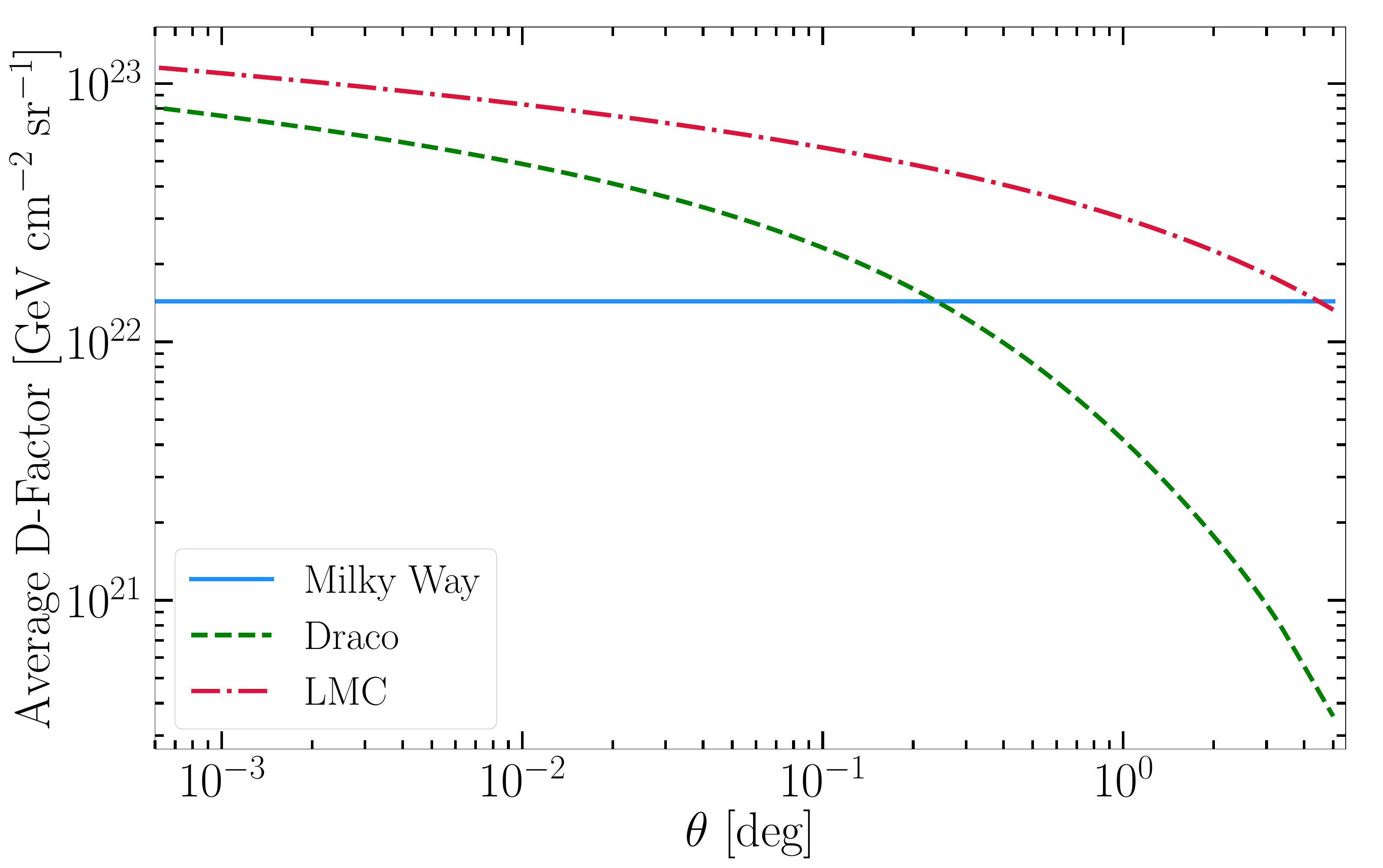}
    \centering
    \caption{Average D-factor (i.e. $D(\Delta\Omega)/\Delta\Omega$) as a function of the angular radius for the Draco dwarf galaxy (dashed green line), the LMC (dash-dotted red) and a circular area of the Milky Way halo centered at the north ecliptic pole (solid blue line).}
    \label{fig:Dfactors}
\end{figure}

Although the ROI is far from the Galactic Center and thus covers a region with lower DM density, when considering the whole area the D-Factor becomes notable, as can be seen in Fig.~\ref{fig:Dfactors}. For this reason, among the targets that we have considered, this is the one that leads to the best sensitivity, as shown in Fig.~\ref{fig:FullSensitivity}, allowing a significant improvement of current bounds on $\gag$.

Another possible strategy to search for ALP decay signals in the halo of the Milky Way is to use a large part of the all-sky survey observations, in particular regions presenting low background emission. Considering, e.g., a region as large as the Euclid Wide Survey, $\sim 15000\, \rm deg^2$ \cite{Euclid:2021icp}, and assuming the noise to scale with the square root of the area, one would gain a factor $\sim 10$ with respect to the case of a Deep Field (i.e., $100\,\rm deg^2$). On the other hand, the all-sky surface brightness sensitivity degrades, by a similar factor, with respect to the Deep Field (see Section~\ref{Sec:spherex}) and thus, considering also the complication of analysing a wider area, we do not expect a significant improvement from an all-sky search. 
A similar conclusion can be obtained for the Galactic Center region, which is a promising target given the potentially large D-factor, but where issues such as the presence of bright regions or dust extinction might be severe.
Even disregarding those potential issues, we find a moderate improvement. For example, considering an angular region of radius $20$ degrees around the Galactic Center, and assuming an NFW density profile, the sensitivity on $g_{a\gamma}$ improves by less than a factor of two with respect to what we obtained for the Deep Field.

Another interesting signature that can be investigated thanks to the wide area covered by SPHEREx is the extragalactic background produced by the ALP decay in DM halos across the Universe. We checked that this spatially uniform contribution leads to weaker constraints than the ones derived from the emission in the Milky Way discussed in this Section. Studying the anisotropies of the ALP extragalactic background can be more promising. Indeed, sensitivities similar to the ones achieved in Fig.~\ref{fig:FullSensitivity} are obtained by the approach discussed in Ref.~\cite{Bernal:2020lkd}, namely searching for the extragalactic ALP signal through the line intensity mapping technique.

\section{Conclusions}\label{Sec:Conclusion}
In this work we have derived projected limits on the axion-photon coupling $\gag$ for ALP masses between 0.5 and 3 eV using the upcoming SPHEREx telescope~\cite{2020SPIE11443E..0IC, SPHEREx:2014bgr,SPHEREx:2016vbo,SPHEREx:2018xfm}. Results are shown in Fig.~\ref{fig:FullSensitivity}. 

ALP models typically feature a coupling to electromagnetism, enabling the possibility of ALPs decaying into two photons. If ALPs make up the cold DM component of the Universe, their decay occurs in the non-relativistic regime inside DM halos, and would produce a nearly monochromatic spectral line at the rest-frame frequency determined by half the ALP mass.

SPHEREx will survey the sky at wavelengths between 0.75 $\mu$m and 5 $\mu$m, meaning that it can detect ALP line signatures for ALP masses between 0.5-3 eV (in the case of nearby targets). SPHEREx is a wide-field telescope, well-suited to search for extended emission like the one coming from DM halos.
We have considered three promising targets: dSph galaxies, LMC, and the halo of the
Milky Way. The latter provides the most competitive bounds, essentially because SPHEREx can integrate the signal over a wide sky area, and thus the D-factor of Eq.~\ref{eq:AxionFlux} becomes very large in the case of the Milky-Way, see Fig.~\ref{fig:Dfactors}.

Our results of Fig.~\ref{fig:FullSensitivity} suggest that current bounds on the axion-photon coupling $\gag$ will be significantly improved by the observations performed by SPHEREx in the forthcoming years and will complement future experiments for masses above 0.5 eV.

\section*{Note added}

In this version of the manuscript we have revised our projected sensitivity to the axion-photon coupling. Our forecasts are based on the surface brightness noise per spectral channel provided in the SPHEREx public repository~\href{https://github.com/SPHEREx/Public-products}{https://github.com/SPHEREx/Public-products} and presented as the sensitivity per spectral channel. In the previous version, we interpreted this quantity as the uncertainty on the intensity integrated over a spectral channel $\delta\nu \cdot I_{\nu}.$ We have then determined that it rather corresponds to the uncertainty of the channel-averaged quantity $\nu \cdot I_{\nu}.$ This correction leads to an improvement of our sensitivities on $g_{a\gamma}$ by a factor $\sqrt{R}=\sqrt{\langle\nu\rangle/\delta\nu}$.

\section*{Acknowledgements}
We thank Connor Hainje, Elena Pinetti and David Hogg for discussions about the SPHEREx surface brightness noise, which prompted us to re-assess our projected sensitivities.
We thank Olivier Dore and James Bock for providing useful information about SPHEREx.
We acknowledge support by the research grant TAsP (Theoretical Astroparticle Physics) funded by Istituto Nazionale di Fisica Nucleare (INFN).
MT and JTC acknowledge the research grant ``Addressing systematic uncertainties in searches for dark matter No.
2022F2843'' funded by MIUR.

\bibliographystyle{bibi}
\bibliography{biblio.bib}

\end{document}